\documentclass[aps,prl,twocolumn,superscriptaddress,groupedaddress]{revtex4}

\usepackage{graphicx}
\usepackage{color}
\usepackage{xspace}
\usepackage{relsize}
\usepackage{aas_macros}
\usepackage{fontenc} 
\usepackage{hyperref}
\usepackage{amsmath}

\usepackage{etoolbox}
\preto\align{\par\nobreak\small\noindent}
\expandafter\preto\csname align*\endcsname{\par\nobreak\small\noindent}
\preto\multline{\par\nobreak\small\noindent}
\expandafter\preto\csname align*\endcsname{\par\nobreak\small\noindent} 
\preto\flalign{\par\nobreak\small\noindent}
\expandafter\preto\csname align*\endcsname{\par\nobreak\small\noindent}
\preto\eqnarray{\par\nobreak\small\noindent}
\expandafter\preto\csname align*\endcsname{\par\nobreak\small\noindent}

\allowdisplaybreaks

\newcommand*{\SONG}{\textsf{SONG}\xspace}
\newcommand*{\intr}{\text{intr}}

\newcommand*{\local}{\text{loc}}

\newcommand*{\sq}{\text{sq}}
\newcommand*{\SN} {\ensuremath{S/N}\xspace}

\newcommand*{\fnl}{\ensuremath{f_{\sub{NL}}}\xspace}
\newcommand*{\fnlbias}{\ensuremath{f_{\sub{NL}}^{\intr}}\xspace}
\newcommand{\sub}[1]{\ensuremath{\text{\smaller #1}}}

\renewcommand{\L}{\ensuremath{\ell}\xspace}

\newcommand{\lmax}{{\ensuremath{\ell_\text{max}}}\xspace}

\newcommand{\lm}{\ensuremath{\ell m}\xspace}

\newcommand{\lone}{\ensuremath{{\ell_1}}\xspace}
\newcommand{\ltwo}{\ensuremath{{\ell_2}}\xspace}
\newcommand{\ltre}{\ensuremath{{\ell_3}}\xspace}
\newcommand{\lmone}{{\ensuremath{\lone m_1}}\xspace}
\newcommand{\lmtwo}{{\ensuremath{\ltwo m_2}}\xspace}
\newcommand{\lmtre}{{\ensuremath{\ltre m_3}}\xspace}
\newcommand*{\sci} [2] {\ensuremath{{#1} \times 10^{#2}}} 

\newcommand*{\avg} [1] {\ensuremath{\left\langle\,{#1}\,\right\rangle}\xspace}

\newcommand*{\eref} [1] {Eq.\ \ref{#1}\xspace}

\newcommand*{\fref} [1] {Figure \ref{#1}\xspace}

\newcommand{\dd}{\textrm{d}}
\newcommand*{\msk}{\\[0.25cm]} 
\newcommand*{\nmsk}{\msk\notag} 
\newcommand{\ldep}{\ensuremath{{\lone\ltwo\ltre}}\xspace}
\newcommand{\F} [2] {\ensuremath{F^{\,({#1}),({#2})}}\xspace} 
\newcommand*{\los}{line-of-sight\xspace}
\newcommand{\threej}[6]{%
\ensuremath{\begin{pmatrix}{#1\!}&{#2\!}&{#3}\\{#4\!}&{#5\!}&{#6}\end{pmatrix}}}

\def\eprinttmp@#1arXiv:#2 [#3]#4@{{\href{http://arxiv.org/abs/#1}{#1}}}
\providecommand{\eprint}[1]{\eprinttmp@#1arXiv: [y]@}

\begin{document}


\title{Impact of polarisation on the intrinsic CMB bispectrum}

\author{Guido W. Pettinari}
\email{g.pettinari@sussex.ac.uk}
\affiliation{Department of Physics \& Astronomy, University of Sussex,
Brighton BN1 9QH, UK}
\affiliation{Institute of Cosmology and Gravitation, University of
Portsmouth, Portsmouth PO1 3FX, UK}

\author{Christian Fidler}
\affiliation{Institute of Cosmology and Gravitation, University of
Portsmouth, Portsmouth PO1 3FX, UK}

\author{Robert Crittenden}
\affiliation{Institute of Cosmology and Gravitation, University of
Portsmouth, Portsmouth PO1 3FX, UK}

\author{Kazuya Koyama}
\affiliation{Institute of Cosmology and Gravitation, University of
Portsmouth, Portsmouth PO1 3FX, UK}

\author{Antony Lewis}
\affiliation{Department of Physics \& Astronomy, University of Sussex,
Brighton BN1 9QH, UK}

\author{David Wands}
\affiliation{Institute of Cosmology and Gravitation, University of
Portsmouth, Portsmouth PO1 3FX, UK}

\date{\today}

\begin{abstract}
We compute the bispectrum induced in the cosmic microwave background (CMB) temperature and polarisation by the evolution of the primordial density perturbations using the second-order Boltzmann code \SONG.
We show that adding polarisation increases the signal-to-noise ratio by a factor four with respect to temperature alone and we estimate the observability of this intrinsic bispectrum and the bias it induces on measurements of primordial non-Gaussianity.
%
When including all physical effects except the late-time non-linear evolution, we find for the intrinsic bispectrum a signal-to-noise of $\SN=3.8,\,2.9,\,1.6$ and $0.5$ for, respectively, an ideal experiment with an angular resolution of $\lmax=3000$, the proposed CMB surveys PRISM and COrE, and Planck's polarised data; the bulk of this signal comes from the $E$-polarisation and from squeezed configurations.
%
We discuss how CMB lensing is expected to reduce these estimates as it suppresses the bispectrum for squeezed configurations and contributes to the noise in the estimator.
We find that the presence of the intrinsic bispectrum will bias a measurement of primordial non-Gaussianity of local type by $\fnlbias=0.66$ for an ideal experiment with $\lmax=3000$.
Finally, we verify the robustness of our results by reproducing the analytical approximation for the squeezed-limit bispectrum in the general polarised case.
\end{abstract}


\maketitle


\paragraph{\textbf{Introduction}}

The three-point function, or bispectrum, of the cosmic microwave background (CMB) is directly linked to non-Gaussian features in the primordial fluctuations from which the CMB evolved \cite{komatsu:2001a, komatsu:2010a, liguori:2010a, yadav:2010a, bartolo:2010a}.
Measuring the CMB bispectrum is therefore equivalent to opening a window to the early Universe.
In particular the temperature maps measured by the Planck CMB survey \cite{planck-collaboration:2013b} provide the most stringent constraint on the amplitude \fnl of primordial non-Gaussianity of the local type \cite{komatsu:2001a, gangui:1994a, verde:2000a}: $\fnl=2.7\,\pm\,5.8 \;$.
Furthermore, the polarised maps, expected from Planck by the end of 2014, will be used to refine the $\fnl$ measurement and reduce the error by approximately a factor two \cite{babich:2004a, komatsu:2005a, yadav:2007a, yadav:2008a, yadav:2010a}.

However not all of the observed non-Gaussianity is of primordial origin.
Indeed, a bispectrum arises in the CMB even for Gaussian initial conditions in the primordial curvature perturbation \cite{lyth:2005b} due to non-linear dynamics such as CMB photons scattering off free electrons and their propagation along a perturbed geodesic in an inhomogeneous Universe.
This \emph{intrinsic bispectrum} is an interesting signal in its own right as it contains information on such processes.
Furthermore, if not correctly estimated and subtracted from the CMB maps, it will provide a bias in the estimate of primordial $\fnl$.

Computing the intrinsic bispectrum requires solving the Einstein and Boltzmann equations up to second order in the cosmological perturbations. These have been studied in great detail \cite{pitrou:2009a, pitrou:2010a, beneke:2010a, naruko:2013a, bartolo:2006a, bartolo:2007a} and approximate solutions have been found in specific limits \cite{bartolo:2004a, bartolo:2004b, boubekeur:2009a, senatore:2009a, senatore:2009b, nitta:2009a}.
In particular, the intrinsic bispectrum can be obtained analytically in the so-called ``squeezed'' limit, where one of the three scales is much larger than the others \cite{creminelli:2004a, creminelli:2011a, bartolo:2012a, lewis:2012a}.
However, for arbitrary configurations, the intrinsic bispectrum has to be computed numerically. 
Numerical convergence is now being reached as the latest numerical codes \cite{huang:2013a, su:2012a, pettinari:2013a, huang:2013b} obtain consistent results.
When considering only the temperature bispectrum, these codes find the bias induced on $\fnl$ by second-order effects to be of order unity, and the intrinsic bispectrum to be unobservable by Planck, its signal-to-noise ratio reaching unity only for an ideal experiment with an angular resolution of $\lmax=3000$.

In this letter, we extend the studies discussed above by including for the first time CMB polarisation, and show that the intrinsic bispectrum signal is enhanced considerably compared to the primordial signals, making it potentially observable in the next generation CMB missions, such as COrE \cite{the-core-collaboration:2011a} and PRISM \cite{prism-collaboration:2013b}.
We also explore the impact that gravitational lensing has on the observability of the intrinsic bispectrum, both by reducing the amplitude of the intrinsic signal and by providing an additional source of noise in the measurement of the bispectra. 

\paragraph{\textbf{Method}}
\label{sec:method}

We recently studied the intrinsic temperature bispectrum \cite{pettinari:2013a, pettinari:2014a} and the $B$-mode polarisation induced from non-linear dynamics \cite{fidler:2014a}. Using the tools developed in the latter paper we extend our bispectrum analysis to include polarisation.  Throughout, we assume the primordial non-Gaussianity to be negligible ($\fnl=0$).

We employ the system of coupled Boltzmann-Einstein equations at second order describing the non-linear evolution of the different species \cite{bartolo:2006a, bartolo:2007a, pitrou:2009a, beneke:2010a, naruko:2013a} 
and work in the Poisson gauge \cite{bertschinger:1996a}. 
The Boltzmann equation consists of a Liouville term, accounting for particle propagation in an inhomogeneous space-time, and a collision term describing particle interactions, i.e. Compton scattering for CMB photons.
We characterise photons by their brightness moments $\Delta_{n}$ with the composite index $n$ including the angular harmonic indices $\lm$ and the polarisation index $X=T,E,B$. The photon intensity is characterised by $X=T$, while $E$ and $B$ characterise the CMB polarisation. 
The Boltzmann equation for $\Delta_n$ reads
\begin{equation}
	\label{eq:boltzmann_compact}
	\dot{\Delta}_{n} \;+\; k\,\Sigma_{nn'}\Delta_{n'} \;+\; \mathcal{M}_{n} 
  \;+\; \mathcal{Q}^L_{n} \;=\; \mathfrak{C}_n \;,
\end{equation}
where $k$ is the Fourier wavevector of the perturbation, and the free-streaming matrix $\Sigma_{nn'}$ encodes the excitation of high-$\ell$ moments over time. We have denoted the terms containing only metric perturbations by $\,\mathcal{M}_{n}$, while $\,\mathcal{Q}^L_{n}\,$ describes the effect of the metric on the photon perturbations, that is the redshift, time-delay and lensing effects.
The collision term
\begin{equation}
	\label{eq:collision_term}
	\mathfrak{C}_n \;=\; -|\dot{\kappa}|\,\left(\;\Delta_{n} \;-\;
  \Gamma_{nn'}\Delta_{n'} \;-\; \mathcal{Q}^C_{n}\,\,\right) \;,
\end{equation}
is proportional to the Compton scattering rate $|\dot{\kappa}|$ and consists of the purely second-order gain and loss terms, and quadratic contributions $\,|\dot{\kappa}|\,\mathcal{Q}^C_n$.
The explicit form and detailed description of all terms can be found in Ref.~\cite{fidler:2014a,pettinari:2013a,beneke:2010a}.

After recombination photons stream freely, so that at conformal time $\eta$ the higher multipoles with $\ell\approx k \,(\eta-\eta_{\text{rec}})$ are excited, making the numerical computation of the photon moments up to today ($\eta_0$) impractical using the full Boltzmann-Einstein equations. In \SONG we instead compute the photon perturbations after recombination using the \los integration \cite{seljak:1996a}:
\begin{equation}
	\label{eq:los_integral}
	\Delta_{n}(\eta_0) \;=\;
    \int_0^{\eta_0}  d\eta \,e^{-\kappa(\eta)} \;
    j_{nn'}(k(\eta_0-\eta))\;\mathcal{S}_{n'}(\eta) \;,
\end{equation}
with the streaming functions $j_{nn'}$ specified in Ref.~\cite{pettinari:2014a, beneke:2011a}, and the line-of-sight source function $\mathcal{S}_{n}$ given by
\begin{equation}
	\label{eq:line_of_sight_sources}
	\mathcal{S}_n \;=\; -\mathcal{M}_{n} \,-\, \mathcal{Q}^L_{n} \,+\,
  |\dot{\kappa}|\left(\Gamma_{nn'}\Delta_{n'}+\mathcal{Q}^C_{n}\right) \;.
\end{equation}
We first solve the full second-order Boltzmann-Einstein hierarchy to build $\mathcal{S}_n$ until the time of recombination, as described in Ref.~\cite{pettinari:2013a,fidler:2014a,pettinari:2014a}, and then compute the \los integral in \eref{eq:los_integral} to obtain the photon perturbations today.

Recently Huang \& Vernizzi (2013) \cite{huang:2013b} clarified the relation between the remapping and second-order Boltzmann approaches to lensing, while Su \& Lim (2014) \cite{su:2014a} developed an alternative non-perturbative treatment of lensing involving a Dyson series. 
These works allow us to identify the lensing and time-delay terms in the second-order equations and remove them from the \los sources, because, at second order, their effect along the \los results in the well-known CMB-lensing bispectrum \cite{lewis:2011a} plus a small residual \cite{huang:2013b}.
We do include the redshift term by using the $\widetilde{\Delta}$ transformation of variables first introduced in Ref.~\cite{huang:2013a} and later generalised to the polarised case in Ref.~\cite{fidler:2014a}:
\begin{align}
  \label{eq:delta_tilde_transformation_multipole_space}
  &
  \begin{aligned}
  \widetilde{\Delta}_{T,\lm} \;=&\; \Delta_{T,\lm} - \frac{1}{2}\,i^{L}
  \left(\begin{array}{cc|c}\L'&\L''&\L \\ m' & m'' & m\end{array}\right) 
  \left(\begin{array}{cc|c}\L'&\L''&\L \\ 0 & 0 &
  0\end{array}\right) \msk
  & \times \Delta_{T,\L'm'}\Delta_{T,\L''m''} \;,
  \end{aligned}
  \nmsk
  &
  \begin{aligned}
  \widetilde{\Delta}_{E,\lm} \;=&\; \Delta_{E,\lm} - i^{L}
  \left(\begin{array}{cc|c}\L'&\L''&\L \\ m' & m'' & m\end{array}\right)
  \left(\begin{array}{cc|c}\L'&\L''&\L \\ 0 & 2 &
  2\end{array}\right) \msk
  & \times \Delta_{T,\L'm'}\Delta_{E,\L''m''} \;,
  \end{aligned}
\end{align}
for even $L=\L-\L'-\L''$. A summation over $\ell',\ell'',m',m''$ is implied and the parentheses symbols represent
the Clebsch-Gordan coefficients.

After using the \los integration to obtain $\Delta_{n}(\eta_0)$, we relate it to the bolometric temperature perturbation $a^X_{\lm}$ \cite{fidler:2014a,pitrou:2014a,pitrou:2010b}, and compute its full-sky bispectrum:
\begin{align}
  B^{XYZ}_\ldep &= \hspace{-4pt}\sum\limits_{m_1m_2m_3}\;
  \threej{\lone}{\ltwo}{\ltre}{m_1}{m_2}{m_3}\;
  \left\langle a^X_{\lmone}\,a^Y_{\lmtwo}\,a^Z_{\lmtre}\right\rangle \notag\msk
  &=\; \threej{\lone}{\ltwo}{\ltre}{0}{0}{0}\;b^{XYZ}_\ldep\;,
  \label{eq:bispectrun_definition}
\end{align}
where $X,Y,Z$ are either $T$ or $E$.
This computation is done in the same way as in our previous work \cite{pettinari:2013a}, where we focussed on the unpolarised scalar contributions to the bispectrum, corresponding to the $m=0$ sources in \eref{eq:line_of_sight_sources}. In this letter, we also include $m\neq0$ contributions up to $|m|=3$ following Ref.~\cite{pettinari:2014a} and find that they are subdominant with respect to the $m=0$ modes as they consist of about $3\%$ of the total signal.


In the calculations below, we have assumed the Planck best-fit $\Lambda$CDM cosmology \cite{planck-collaboration:2013a} where $h=0.678$, $\Omega_b=0.0483$, $\Omega_{\text{cdm}}=0.259$, $\Omega_\Lambda=0.693$, $A_s=\sci{2.214}{-9}$, $n_s=0.961$, $N_\text{eff}=3.04$, $\kappa_\text{reio}=0.095$. We also assume adiabatic initial conditions with a vanishing primordial tensor-to-scalar ratio ($r=0$).

\paragraph{\textbf{Polarisation impact}}
\label{sec:analytical}

\begin{figure}[t]
  \centering
    \includegraphics[width=0.5\textwidth]{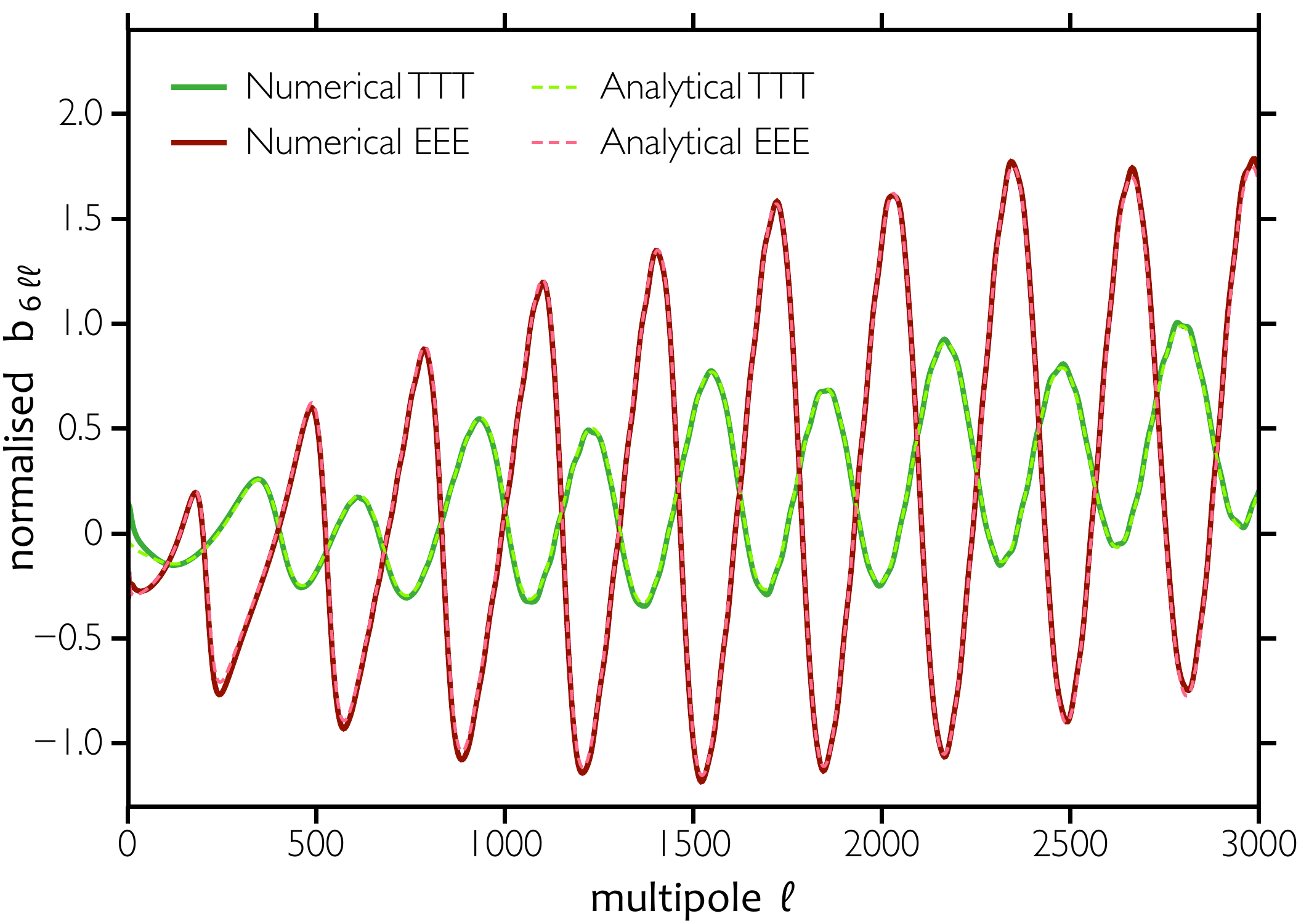}
  \caption{
  Numerical intrinsic bispectrum produced by \SONG, together with the squeezed-limit approximation in \eref{eq:squeezed_limit_bispectrum}, with $\lone=6$ and $\ltwo=\ltre=\ell$, for $TTT$ and $EEE$. We normalise the curves with respect to the squeezed limit for the local-type bispectrum \cite{gangui:1994a, komatsu:2001a} with $\fnl=5$, that is $6\,C_{\lone}^{X\zeta}\left(\,C^{YZ}_{\ltwo}\,+\,C^{YZ}_{\ltre}\,\right)\,$.
  }
  \label{fig:squeezed_limit_bispectrum}
\end{figure}

For squeezed configurations ($\lone\ll\ltwo,\,\ltre$), where the long-wavelength mode is within the horizon today but was not at recombination ($\lone\ll200$), the intrinsic bispectrum is known analytically.
In this case, the super-horizon curvature perturbation at recombination, $\zeta$, acts as a perturbation to the background curvature that dilates the observed angular scale of the small-scale CMB anisotropies.
Large and small scales are thus correlated and a squeezed intrinsic bispectrum arises that is proportional to the correlation between the large-scale CMB anisotropies and $\zeta$.
%
%
In multipole space, a dilation corresponds to a sideways shift in $\L$, so that the more sharply peaked the small-scale power spectrum, the bigger the change in power.
%
As a result, the intrinsic bispectrum in the squeezed limit is proportional to the derivative of the small-scale power spectrum \cite{creminelli:2004a, creminelli:2011a, bartolo:2012a, lewis:2012a}:
\begin{align}
	\label{eq:squeezed_limit_bispectrum}
	b^{\sq,XYZ}_\ldep \;=\; 
    &-\frac{1}{2} \; C_{\lone}^{X\zeta} \;\left[\,%
    \frac{\dd\,(\ltwo^2\,C^{YZ}_{\ltwo})}{\ltwo\,\dd\,\ltwo} \,+\,%
    \frac{\dd\,(\ltre^2\,C^{YZ}_{\ltre})}{\ltre\,\dd\,\ltre}%
    \,\right]
    \msk
    &+C^{XT}_{\lone} \;\left[\,%
    \delta_{ZT}\,C^{YT}_{\ltwo} \,+\, \delta_{YT}\,C^{ZT}_{\ltre}
    \,\right]
    \;, \notag
\end{align}
where $X,Y,Z$ are either $T$ or $E$ and the angular power spectra are defined by
$  \avg{a^X_{\ell m}\,a^{Y*}_{\ell' m'}} \;\equiv\;
  C^{XY}_\ell\,\delta_{\ell\ell'}\,\delta_{mm'} \;.$
The second line in \eref{eq:squeezed_limit_bispectrum} represents the subdominant effect known as redshift modulation \cite{lewis:2012a}.
%
%
In \fref{fig:squeezed_limit_bispectrum}, we show that \SONG's numerical bispectra match the analytical approximation in the squeezed limit at percent-level precision.



In \SONG we truncate the line-of-sight integration in \eref{eq:los_integral} at recombination and thus neglect the second-order scattering sources at reionisation. Their computation is challenging as it involves summations over high-$\ell$ multipoles at late times. In any case, a second-order treatment would still be insufficient, as non-linear effects are relevant at the time of reionisation.
We do however include reionisation at the background and linear level.
The squeezed formula of \eref{eq:squeezed_limit_bispectrum} works at the same level since $\zeta$ is defined to be the value of the curvature perturbation at recombination, which is the source of the small-scale perturbations being modulated; hence the match with \SONG for squeezed shapes shown in \fref{fig:squeezed_limit_bispectrum}.
The linear temperature anisotropies observed today are sourced by density and velocity perturbations at recombination. Since these are out-of-phase in $\ell$ space, the resulting acoustic peaks are blurred.  On the other hand, the peaks of the $E$ polarisation spectrum are sharper as their only source is the quadrupole induced by Compton scattering. 
It follows that the logarithmic derivatives in \eref{eq:squeezed_limit_bispectrum} normalised to $C^{YZ}_{\L}$ will be larger for polarisation than for temperature. This enhancement is about a factor $2.5$ in magnitude across the whole $\ell$-range (\fref{fig:squeezed_limit_bispectrum}) and leads to a larger signal-to-noise ratio for the polarised bispectra.
Note, however, that the temperature bispectrum, $TTT$, is still much larger than the polarisation one, $EEE$, as $C_{\lone}^{T\zeta}\approx100\,C_{\lone}^{E\zeta}$ for $\lone<200$ \cite{lewis:2012a}.


In order to quantify the observability of the intrinsic bispectrum and its bias on a primordial measurement of \fnl, we build the Fisher matrix element in the general polarised case as \cite{babich:2004a, yadav:2008a, komatsu:2001a, lewis:2011a} 
\begin{align}
  \label{eq:fisher_definition}
  &F^{(i),(j)} \;=\;
    \sum\limits_{ABC,XYZ}\;\sum\limits_{2\leq\lone\leq\ltwo\leq\ltre}^{\lmax}
    \frac{1}{\Delta_\ldep}\; \nmsk
    &\;
    \times\;B^{(i),ABC}_\ldep \;
    \left(\widetilde{C}_\text{tot}^{-1}\right)^{AX}_\lone\;
    \left(\widetilde{C}_\text{tot}^{-1}\right)^{BY}_\ltwo\;
    \left(\widetilde{C}_\text{tot}^{-1}\right)^{CZ}_\ltre\;
    B^{(j),XYZ}_\ldep \;,
\end{align}
where $\,\Delta_\ldep = 1,2,6\,$ for triangles with no, two or three equal sides, $\,\lmax$ is limited by the finite angular resolution of the survey and $\,\widetilde{C}_\text{tot}^{XY}$ is given by the sum of the lensed spectrum and a noise term to account for the sensitivity of the survey \cite{pogosian:2005a, knox:1995a}.
The first sum involves all possible pairs of the eight bispectra ($TTT$, $TTE$, $TET$, $ETT$, $EET$, $ETE$, $TEE$, $EEE$), while the product of three $\widetilde{C}^{-1}_\text{tot}$ represents their covariance.
The observability of the intrinsic bispectrum is quantified by its signal-to-noise ratio $\SN=\sqrt{F^{\intr,\intr}}\,$, while the bias it induces on a measurement of local-type $\fnl$ is $\,\fnlbias=F^{\local,\intr}/F^{\local,\local}\,$.

\begin{figure}[t]
  \centering
    \includegraphics[width=0.5\textwidth]{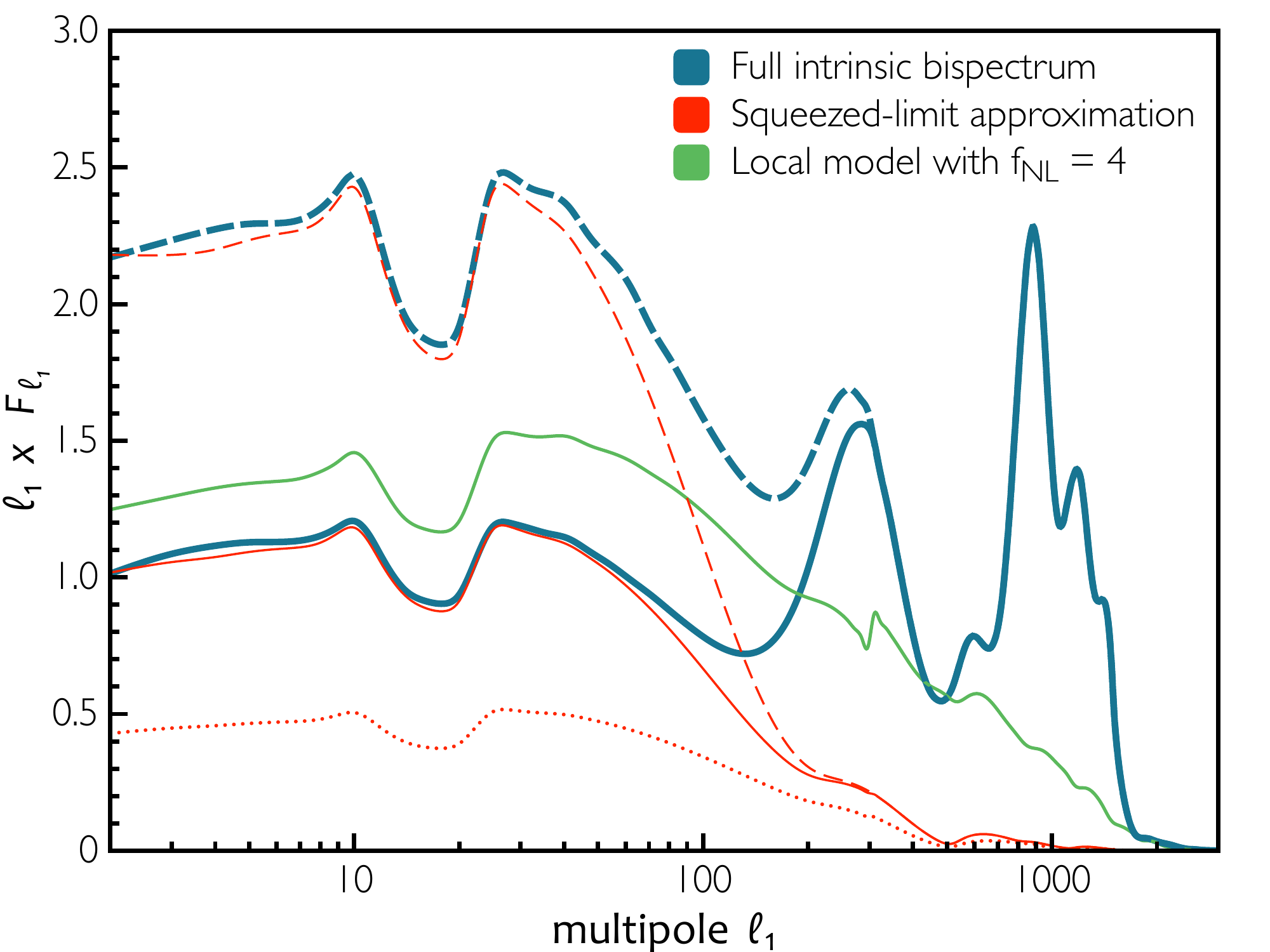}
  \caption{
  Contributions (per $\ln\lone$) to the inverse variance, $\,\F{i}{i}\,$, of the intrinsic bispectrum as a function of the smallest multipole in the sum, $\lone$, considering an ideal experiment with $\lmax=3000$. In this plot, the area below each bispectrum curve is proportional to the square of its signal-to-noise ratio.
  The full bispectrum computed with \SONG (thick-blue lines) exhibits distinctive peaks on subhorizon scales ($\lone>100$), while on larger scales it tracks the the squeezed-limit approximation (thin red lines). 
The dashed lines do not include lensing variance, which reduces the signal-to-noise only on superhorizon scales; the lensed bispectrum in the squeezed limit is shown by the dotted line.  
  For reference, we also include the curve expected from a local-type bispectrum with $\fnl=4$ (green solid line).
  }
  \label{fig:fisher_l1}
\end{figure}

In \fref{fig:fisher_l1} we show the signal-to-noise as a function of $\lone$, the smallest multipole in the Fisher matrix sum, for an ideal CMB survey with a resolution of $\lmax=3000$. 
On superhorizon scales ($\lone\ll200$), the intrinsic bispectrum computed by \SONG agrees well with the squeezed-limit formula, as expected.
The subhorizon effects computed by \SONG become important for $\lone>100$ and give rise to several acoustic peaks.
The signal associated to the subhorizon peaks is given by the square root of the area below the curve, and amounts to $\SN=1.6$ and $2.7$ for $\lmax = 3000$ and $4000$, respectively; most of this signal comes from squeezed triangles.
Note that these effects cannot be treated in the analytical approximation in \eref{eq:squeezed_limit_bispectrum} and hence need to be computed using a full second-order code like \SONG. Furthermore, we find that the relative importance of the subhorizon effects increases with $\lmax$.

\paragraph{\textbf{Lensing effects}}

The Fisher matrix estimator of \eref{eq:fisher_definition} is optimal only under the assumption of a nearly Gaussian CMB. However, the gravitational lensing of CMB photons generates a non-Gaussian signal that must be accounted for in the covariance matrix; not doing so would overestimate the significance for a detection of the intrinsic bispectrum \cite{lewis:2011a, hanson:2009a}.
We account for this \emph{lensing variance} in the estimator following the analytic approach of Ref.~\cite{lewis:2011a}, which is valid for squeezed configurations.

We find that lensing variance degrades the intrinsic signal from $\lone<200$ by approximately a factor  $\sqrt{2}$, while leaving the signal from smaller scales unaltered, as can be seen by comparing the dashed and solid curves in \fref{fig:fisher_l1}.
The reason is that most of the signal below $\lone<200$ comes the squeezed configurations described by the analytical formula in \eref{eq:squeezed_limit_bispectrum}, which are highly degenerate with the isotropic part of lensing (i.e., convergence) \cite{creminelli:2004a, creminelli:2011a, lewis:2012a}.
As a result, the added noise from lensing convergence significantly reduces our ability to detect the intrinsic bispectrum for $\lone < 200$.
On the other hand, we find that the intrinsic signal does not correlate significantly with convergence or shear modes on smaller scales and is thus not affected by lensing variance.
After correcting for lensing variance, the subhorizon effects constitute about $40\%$ and $50\%$ of the intrinsic signal squared for $\lmax=3000$ and $4000$, respectively, and a larger fraction for higher resolutions.

In addition to the lensing-induced variance, two other effects of gravitational lensing may affect the measurement of the intrinsic bispectrum.
First, the correlation between the photon intensity and the lensing deflection angle
results in the emergence of a CMB-lensing bispectrum \cite{spergel:1999a, goldberg:1999a, seljak:1999a} which was recently detected by the Planck experiment \cite{planck-collaboration:2013b}, and corresponds to the lensing terms we dropped in the \los integration.
The isotropic part of the CMB-lensing bispectrum is known to be degenerate with the intrinsic bispectrum \cite{creminelli:2004a, creminelli:2011a, lewis:2012a} in the squeezed limit and might therefore contaminate a measurement of the latter.
To quantify this effect we include the amplitude of the CMB-lensing bispectrum in our Fisher matrix and marginalise over it. We find that, in the full polarised case and including lensing variance, the intrinsic and lensing bispectra have a correlation of $0.6\%$ and, therefore, the intrinsic $\SN$ is degraded only by about $0.002\%$.
These numbers suggest that the CMB-lensing bispectrum is different enough from the intrinsic bispectrum to allow a clear separation of the effects, in analogy to the case of the local \fnl template \cite{lewis:2011a, planck-collaboration:2013b}.
Note that this separation cannot be used to reduce the impact of lensing variance, as the signal cannot be used to reduce the noise in the estimator.

Secondly, CMB lensing distorts the observed shape of the intrinsic bispectrum in a non-perturbative way. In the squeezed limit, the lensed bispectrum is obtained by substituting the power spectrum $C_\ell^{YZ}$ in \eref{eq:squeezed_limit_bispectrum} with its lensed counterpart, $\widetilde{C}_\ell^{YZ}$ \cite{lewis:2012a, lewis:2011a}. This results in a smaller bispectrum as the derivatives in \eref{eq:squeezed_limit_bispectrum} now act on a smoother function. The $\SN$ from squeezed configurations is consequently reduced by a factor of approximately $\,\sqrt{2}\,$ for $\lmax=3000$, as can be seen by comparing the solid and dotted red curves in \fref{fig:fisher_l1}.
However, this suppression is only valid in the squeezed case; for arbitrary configurations one has to resort to a more general approach \cite{cooray:2008a, hanson:2009a, pearson:2012a}, which we will address in future work.  Here we focus on the unlensed intrinsic bispectra.

\paragraph{\textbf{Observational implications}}

\begin{figure}[t]
  \centering
    \includegraphics[width=0.5\textwidth]{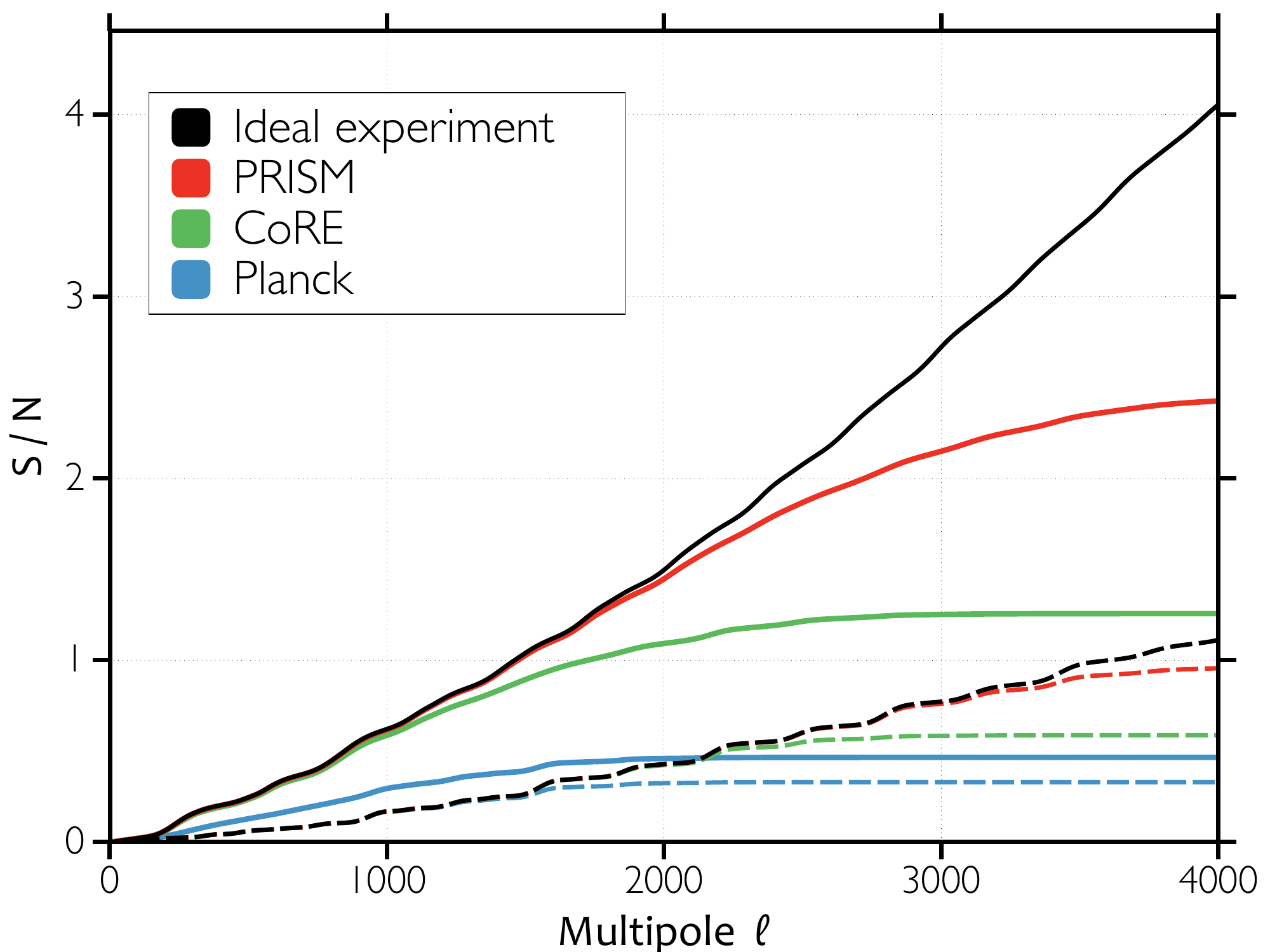}
  \caption{
  Signal-to-noise ratio of the intrinsic bispectrum for the four experiments described in the text as a function of the maximum resolution \lmax, including lensing variance and excluding the lensing of the bispectrum. The solid curves include all eight temperature and polarisation bispectra, the dashed ones only temperature.
  }
  \label{fig:signal_to_noise}
\end{figure}

In \fref{fig:signal_to_noise} we present the signal-to-noise ratio of the intrinsic bispectrum for four experiments as a function of the maximum resolution \lmax. We consider an ideal experiment, the proposed CMB surveys PRISM \cite{prism-collaboration:2013b} and COrE \cite{the-core-collaboration:2011a}, and Planck polarisation data.
We only employ frequencies between $100$ and $300$ GHz, where the CMB signal peaks, and assume full-sky observations.
Through numerical convergence tests for the key numerical parameters, we have ensured these results are stable at the percent-level.

\fref{fig:signal_to_noise} shows that the angular resolution of the CMB survey strongly affects the detectability of the intrinsic bispectrum.
With a resolution of $\lmax=3000$, an ideal experiment would observe the intrinsic bispectrum at the $2.7\sigma$ level, while PRISM, COrE and Planck polarised data would yield $2.1\sigma,1.3\sigma,0.46\sigma$, respectively.
When we account for the lensing of the bispectrum in the squeezed regime via the analytical formula in \eref{eq:squeezed_limit_bispectrum}, these numbers reduce to $2.1\sigma,1.8\sigma,1.1\sigma,0.41\sigma$.

Another important feature of \fref{fig:signal_to_noise} is that most of the signal in the intrinsic bispectrum comes from $E$-polarisation rather than temperature, despite the fact that only a $10\%$ fraction of the CMB anisotropies are polarised.
The reason is twofold. 
First, the dilation effect that generates the intrinsic bispectrum on squeezed scales is $\sim\!2.5$ times more efficient for polarisation than for temperature, as shown in \fref{fig:squeezed_limit_bispectrum}.
Secondly, as long as the instrumental noise is low enough, both temperature and polarisation are sample-variance limited, so that the polarisation bispectrum variance is also suppressed compared to that of the temperature. 
In principle, the same argument applies to $B$-polarisation. The intrinsic $B$-signal, however, is sourced by non-scalar sources that are geometrically suppressed \cite{fidler:2014a} making it smaller than the $E$-signal and thus likely to be dominated by lensing \cite{zaldarriaga:1998c, lewis:2006a} and instrumental noise. 

We find the bias to the local-type \fnl for an ideal experiment with resolution $\,\lmax=3000\,$ to be $\,\fnlbias=1.33,1.50,1.51\,$ for temperature, polarisation and the two probes combined, respectively.
Including lensing variance reduces the bias to $\fnlbias=0.95,0.61,0.66$. This suppression is due to the intrinsic bispectrum being affected by lensing variance more than the local template.
The bias is further reduced by varying the experimental setup: for PRISM, CoRE and Planck polarised data we find $\fnlbias=0.58,0.45,0.37$, respectively, considering lensing variance and both temperature and polarisation.


\paragraph{\textbf{Conclusions}}

Including polarisation is crucial to extract all the information contained in the CMB. In this letter we have extended previous analyses of the intrinsic bispectrum \cite{pettinari:2013a, huang:2013b, su:2012a} and shown that it is particularly sensitive to polarisation due to the sharp acoustic peaks in the $E$-mode power spectrum. 
Using a Fisher matrix approach, we showed that the eight combined bispectra generate a signal-to-noise ratio four times larger with respect to the temperature-only case, making the signal potentially observable at the $2\sigma$ level in future high-resolution missions, such as PRISM \cite{prism-collaboration:2013b} or an improved version of COrE \cite{the-core-collaboration:2011a}.
Despite the enhancement of the intrinsic signal, we still find its contamination to the local-type primordial $\fnl$ to be comparable to the unpolarised case.

For squeezed configurations, the gravitational lensing of CMB photons limits the possibility of observing the intrinsic bispectrum by adding extra variance and by reducing its observed amplitude. These effects combine to reduce the signal-to-noise by a factor of two.  However, on subhorizon scales, where full second-order codes such as \SONG are crucial, the added variance does not limit the detectability.

Here we have included the effects of reionisation and the lensing of the bispectrum only in an approximate way, focussing on their impact in the squeezed limit.  Reionisation could lead to additional intrinsic contributions in a full second-order treatment, while lensing could affect the signal on subhorizon scales.  We will examine these questions in future work.  

\paragraph{\textbf{Acknowledgements}}
GWP and AL acknowledge support by the UK STFC grant ST/I000976/1; CF, RC, KK and DW are supported by STFC grants ST/K00090/1 and ST/L005573/1.
The research leading to these results has received funding from the European Research Council under the European Union's Seventh Framework Programme (FP/2007-2013) / ERC Grant Agreement No. [616170].

\bibliography{my_bibliography}

\end{document}